\documentclass[12pt]{article}
\usepackage[utf8]{inputenc}
\usepackage{amsmath,amssymb,amsthm}
\usepackage{graphicx,epsf, color}
\usepackage{enumerate}
\usepackage{natbib}
\usepackage{forest}
\usepackage{enumitem}
\usepackage{float}
\usepackage{url} 
\usepackage{times}
\usepackage[ruled,noend]{algorithm2e}
\usepackage{graphics}
\usepackage{booktabs}
\usepackage{xr}
\usepackage[colorlinks=true, linkcolor=blue, citecolor=blue, filecolor=blue, urlcolor=blue]{hyperref}
\usepackage{setspace}
\usepackage{caption}
\usepackage{subcaption}

\newcommand{\blind}{1}

\usepackage[margin=1in]{geometry}

\catcode`\^ = 13 \def^#1{\sp{#1}{}}

\definecolor{Pink}{rgb}{1.0, 0.5, 0.5}
\definecolor{Maroon}{rgb}{0.8, 0.0, 0.0}

\def\boxit#1{\vbox{\hrule\hbox{\vrule\kern6pt\vbox{\kern6pt#1\kern6pt}\kern6pt\vrule}\hrule}}

\newtheorem{theorem}{Theorem}
\newtheorem{lemma}[theorem]{Lemma}

\newcounter{remarkcounter}

\newenvironment{remark}[1][Remark \arabic{remarkcounter}.]{
    \refstepcounter{remarkcounter} 
    \begin{trivlist}
    \item[\hskip \labelsep {\itshape #1}]
}{
    \end{trivlist}
}

\newcommand{\bt}{\mbox{\bf t}}

\newcommand{\bzero}{\mbox{\bf 0}}

\newcommand{\bbeta}{\mbox{\boldmath $\beta$}}
\newcommand{\beeta}{\mbox{\boldmath $\eta$}}

\newcommand{\btheta}{\mbox{\boldmath $\theta$}}

\newcommand{\bGamma}{\mbox{\boldmath $\Gamma$}}

\newcommand{\bSigma}{\mbox{\boldmath $\Sigma$}}

\newcommand{\bpsi}{\mbox{\boldmath $\psi$}}

\newcommand{\bbE}{\mathbb{E}}

\newcommand{\var}{\mathrm{Var}}

\newcommand{\tr}{\mathrm{tr}}

\def\t{^\top}

\def\beqn{\begin{eqnarray}}
\def\eeqn{\end{eqnarray}}
\def\beqns{\begin{eqnarray*}}
\def\eeqns{\end{eqnarray*}}

\def\0{{\bf 0}}

\def\h{{\bf h}}

\def\m{{\bf m}}

\def\t{{\bf t}}

\def\s{{\bf s}}

\def\x{{\bf x}}

\def\1{{\bf 1}}

\def\trans{^{\rm T}}

\def\t1trans{^{t+1\rm T}}

\begin{document}

\def\spacingset#1{\renewcommand{\baselinestretch}%
{#1}\small\normalsize}
\spacingset{1}



\title{Surrogate-Powered Inference: Regularization and Adaptivity}


\if0\blind
{
  \author{}\date{}
  \maketitle
} \fi

\if1\blind
{
  \author{Jianmin Chen$^1$, Huiyuan Wang$^1$, Thomas Lumley$^2$, Xiaowu Dai$^3$, Yong Chen$^{1}$\thanks{Corresponding author; ychen123@upenn.edu}\\    
$^1$\textit{Department of Biostatistics, Epidemiology and Informatics,}\\ \textit{University of Pennsylvania}\\
$^2$\textit{Department of Statistics, University of Auckland}\\
$^3$\textit{Department of Statistics and Data Science, University of California, Los Angeles}
}
}
\fi

\maketitle

\normalsize

\begin{abstract}

\noindent
High-quality labeled data are essential for reliable statistical inference, but are often limited by validation costs. While surrogate labels provide cost-effective alternatives, their noise can introduce non-negligible bias. To address this challenge, we propose the surrogate-powered inference (SPI) toolbox, a unified framework that leverages both the validity of high-quality labels and the abundance of surrogates to enable reliable statistical inference. SPI comprises three progressively enhanced versions. Base-SPI integrates validated labels and surrogates through augmentation to improve estimation efficiency. SPI+ incorporates regularized regression to safely handle multiple surrogates, preventing performance degradation due to error accumulation. SPI++ further optimizes efficiency under limited validation budgets through an adaptive, multiwave labeling procedure that prioritizes informative subjects for labeling. Compared to traditional methods, SPI substantially reduces the estimation error and increases the power in risk factor identification. These results demonstrate the value of SPI in improving the reproducibility. Theoretical guarantees and extensive simulation studies further illustrate the properties of our approach.
\end{abstract}

\noindent%

\doublespacing

\clearpage


\section{Introduction}

High-quality labels are essential for ensuring the reliability and reproducibility of data-driven decision-making, yet access to such infallible data is often prohibitively expensive. One prominent example arises in sentiment analysis, where annotating text typically requires a labor-intensive and time-consuming process. In practice, such annotations are often constrained by limited resources, making accurate labels available only for a small subset of objects~\citep{overhage2020physician}.
Subsequent analyses supervised by these validated labels can be reliable, but their efficiency is largely limited by the small sample size.
To address the scarcity of high-quality labels, practitioners have shifted towards using surrogate labels: abundant but noisier alternatives that are informative of the validated label.
While the high-throughput surrogates have the advantage of a much larger sample size, using such noisy labels as the ground truth can introduce non-negligible bias in downstream tasks ~\citep{kern2013accuracy, dubel2023impact}. 

\begin{figure}[tb]
    \centering
    \includegraphics[width=.6\linewidth]{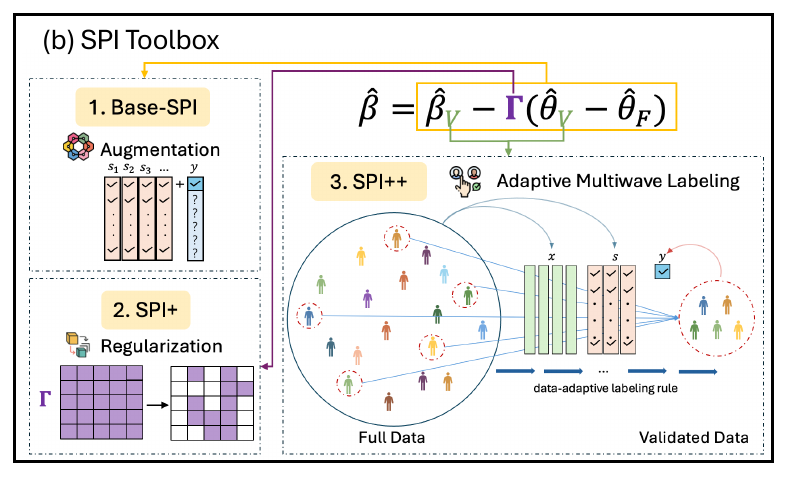}
    \caption{Surrogate-powered inference (SPI) framework. 
    In the figure, $\hat\bbeta_{\mathcal{V}}$ is the estimated parameter of interest on the validated data, $\hat\btheta_{\mathcal{V}}$ and $\hat\btheta_{\mathcal{F}}$ are the estimated nuisance parameters on the validated data and the full data, $\hat\bbeta$ is the final estimate, and $\bGamma$ is a weighting matrix. From Base-SPI to SPI+ and SPI++, the toolbox combines augmentation, regularization, and adaptive multiwave labeling for better performance.}
    \label{fig:spi}
\end{figure}

To harness both the validity of high-quality labels and the rich information contained in surrogates, we introduce \textbf{s}urrogate-\textbf{p}owered \textbf{i}nference (SPI), a unified framework for reliable statistical inference. 
The SPI toolbox consists of three variants: the Base-SPI and its variants SPI+ and SPI++, as illustrated in Figure~\ref{fig:spi} (b).

The Base-SPI builds upon works of augmentation-based semi-supervised inference \citep{sohn2020fixmatch, tong2020augmented, xie2020unsupervised}, and proposes a novel surrogate-powered framework that extends to general $m$-estimators. Specifically, Base-SPI adopts an augmented estimation approach that combines $m$-estimators for the target parameter and surrogate-related nuisance parameters through a weighted adjustment, which yields improved efficiency while preserving asymptotic unbiasedness. However, as the dimension of nuisance parameters grows with the inclusion of more surrogates, Base-SPI may experience performance degradation~\citep{lu2024leveraging}.

To optimally combine high-throughput surrogates and limited human-annotated labels, we propose SPI+, which leverages regularization to mitigate error accumulation from high-dimensional surrogate information and ensures that integrating surrogates does not degrade performance. SPI+ safely incorporates multiple surrogates of varying quality by pursuing sparsity within the augmentation process. As surrogates often approximate the same underlying label, SPI+ directly embeds a regularized regression step to reduce redundancy and control estimation error. 
Inheriting the efficiency gains of Base-SPI, SPI+ further satisfies the no-harm principle: combining surrogate information should not harm estimation performance \citep{raise2024no}.

Building on SPI+, we further propose SPI++, which improves labeling efficiency by adaptively selecting validation subjects. Rather than labeling subjects uniformly at random, SPI++ prioritizes those expected to minimize model uncertainty. This enhances estimation efficiency under labeling constraints, particularly when the validation process can be actively controlled.
Such guided label inquiry is often overlooked, and the selection rule seldom accounts for both validated labels and surrogates.
SPI++ bridges this gap by formulating a combinatorial problem that jointly performs augmented estimation and labeling to improve efficiency, and further adopts a multiwave labeling strategy with an actively updated labeling rule.

These contributions address the dual challenges of harmful integration of surrogates and inefficient labeling, making SPI a powerful inference framework with broad applications. It leverages the interactive nature of label review to enable guided adaptive labeling, maximizing the use of a limited budget while effectively utilizing noisy data without negative effects.

\subsection{Related Work and Methodological Contributions}

SPI relates to the existing literature on estimation with many unlabeled observations and a small number of labeled observations, which primarily focuses on improving efficiency in semiparametric models \citep{pepe1992inference, lawless1999semiparametric, zhu2005semi, wasserman2007statistical, zhang2019semi, kallus2025role}.  In particular, \citet{song2024general} studied general $m$-estimation by applying a projection-based correction to the classical loss function, using low-order basis expansions of the covariates to recover efficiency. While SPI continues in this general direction, it departs in two key ways: it is specifically designed to integrate surrogate outcomes, rather than the covariate-driven structure. Moreover, SPI extends beyond single-stage estimation by incorporating regularization for high-dimensional surrogate information adjustment and an adaptive labeling strategy to optimize sample efficiency under labeling constraints.

The augmented estimation strategy in Base-SPI for integrating noisy and validated data can be traced back to early works by \citet{robins1994estimation}, \citet{chen2000unified}, and \citet{chen2008improving}, in which noise can arise from covariates, outcomes, or both. Focusing on noisy outcomes, 
SPI flexibly combines surrogates from different sources without imposing assumptions on the surrogate. 
The use of prediction-based surrogates has recently gained attention in the prediction-powered inference framework (PPI)~\citep{angelopoulos2023prediction,zrnic2024cross}. While both SPI and PPI are augmentation-based approaches, PPI may suffer performance degradation when the predictive model has substantial errors, leading to confidence intervals that are significantly wider than those from classical methods. In contrast, SPI enjoys the no-harm principle in the sense that incorporating surrogates does not degrade inference performance. Notably, this no-harm principle of SPI is analogous to the critical goal of avoiding negative data-integration in building prediction models with noisy data~\citep{li2019towards}.

SPI also connects to informative sampling, particularly when sampling probabilities are designed to minimize the variance of an estimator. 
\citet{ma2014statistical} proposed a leverage-based sampling method in least-squares problems, while \citet{wang2018optimal} derived an optimal sampling rule under logistic regression. Both methods assume the true outcomes are available for all subjects, and the latter procedure was extended to the missing label scenario in \citet{zhang2021optimal}. With a single surrogate, \citet{liu2022sat} modified the sampling rule in \citet{wang2018optimal} by replacing true labels with the surrogate. While these methods rely on one-time estimated sampling rules, \citet{chen2020optimal} and  \citet{zrnic2024active} considered iteratively refining sampling rules with newly obtained labels. 

SPI formulates a combinatorial problem that jointly performs augmentation and sampling, leveraging both validated and surrogate labels within a unified pipeline. While recent works~\citep{chen2020optimal, wang2023maximin} have explored related ideas, SPI goes further by incorporating adaptive sampling and the no-harm principle, ensuring robust and reliable inference.

\subsection{Organization of the Paper}

Section~\ref{sec:estimator} presents the general SPI framework. In Section~\ref{sec:regaug}, we incorporate a regularization technique into SPI+ to control estimation error and establish the main theoretical results. To further enhance efficiency, we propose SPI++ with a multiwave adaptive labeling design in Section~\ref{sec:sampling}. Numerical studies with synthetic data are conducted in Section~\ref{sec:simu}. Further discussions are provided in Section~\ref{sec:diss}. We have also developed an \textit{R} package for the SPI toolbox.

\section{SPI: Surrogate-Powered Inference}
\label{sec:estimator}


\subsection{Problem Setup}

Let $y\in\mathbb{R}$ be the validated label, $\x=(x_1, \dots, x_p)^{\trans} \in\mathbb{R}^p$ be the $p$-dimensional feature vector, and $\s=(s^{(1)}, \dots, s^{(K)})^{\trans}\in\mathbb{R}^K$ be the collection of $K$ surrogates.
Suppose the full dataset $\mathcal{F}$ contains $N$ independent copies of $(\x,\s)$, and the validation set $\mathcal{V}$ contains $n$ observations of $(\x,\s, y)$, which are selected for labeling through independent Poisson sampling. We assume that the selection probability depends on the fully observed data $(\x, \s)$, and define the selection rule as $\rho(\x,\s)=P(R=1\mid \x,\s)$, where $R=1$ indicates that a subject is selected for outcome labeling, and $R=0$ otherwise. 
We consider the selection design satisfying a pre-specified constraint $\bbE\{\rho(\x,\s)\}\leq\pi$ ($0<\pi<1$), so that the expected number of validated subjects is no more than $\pi N$, where $\pi$ is usually controlled by the total validation budget and the per-subject labeling cost. 
Throughout the discussion, we assume that there exists a positive constant $\varepsilon$ such that $P(R=1\mid \x,\s) \geq \varepsilon >0$ for all $(\x,\s)$. 

We are interested in the relationship between the validated label $y$ and the covariates $\x$, specified by the $p$-dimensional parameter $\bbeta_0$, such that
\begin{equation*}
    \bbE\{\m(\bbeta_0;y,\x)\} = \bzero_p
\end{equation*}
for some function $\m(\cdot)\in\mathbb{R}^p$. For example, $\m(\cdot)$ can be the score function under a generalized linear model (GLM). Given a known selection rule $\rho(\x,\s)$, an $m$-estimator $\hat\bbeta_{\mathcal{V}}$ can be obtained from the independent samples in the validation set by solving the following estimating equation 
\begin{equation}\label{eq:est-beta}
    \frac{1}{N}\sum_{i\in \mathcal{V}} \frac{1}{\rho(\x_i,\s_i)}\m(\bbeta;y_i,\x_i)
    = \bzero_p.
\end{equation}
Although the resulting estimator is asymptotically unbiased under regularity conditions, the estimation is restricted to the small labeled dataset $\mathcal{V}$, leading to substantial information loss. Moreover, the validation set is commonly drawn as a random sample from $\mathcal{F}$ in practice, and the selection strategy is not designed in particular to improve efficiency. 

The main objective of this work is to boost the efficiency of statistical inference through a combination of surrogate integration and adaptive labeling.
We consider two guiding principles: A-optimality and no-harm integration. The A-optimality criterion considers an estimator more efficient if the sum of its component variances, $\sum_{j=1}^p\var(\hat\bbeta_j)$, is smaller, which is equivalent to the trace of the variance-covariance matrix~\citep{kiefer1959optimum}.
The no-harm principle requires that the proposed estimator be at least as efficient as the baseline estimator $\hat\bbeta_{\mathcal{V}}$, which does not incorporate surrogate information~\citep[e.g.,][]{lin2013agnostic}. Throughout the following discussion, we refer to variance minimization as minimizing its trace.

\subsection{Method}

We now introduce the surrogate-powered inference (SPI) framework. Let $\btheta_0^{(k)}\in\mathbb{R}^d$ be a nuisance parameter which captures the relationship between $s^{(k)}$ and $\x$ such that 
\begin{equation*}
    \bbE\{\h(\btheta_0^{(k)};s^{(k)}, \x)\} = \bzero_d, \quad k=1, \dots, K,
\end{equation*}
for some function $\h(\cdot)\in\mathbb{R}^d$. For each of the $K$ surrogate outcomes $s^{(k)}$, we obtain the $m$-estimators of $\btheta_0^{(k)}$ from either $\mathcal{V}$ or $\mathcal{F}$ as the solution to the following estimating equations: 
\begin{equation}\label{eq:est-theta}
    \hat\btheta_{\mathcal{V}}^{(k)}:\frac{1}{N}\sum_{i\in\mathcal{V}} \frac{1}{\rho(\x_i,\s_i)}\h(\btheta^{(k)};s_i^{(k)},\x_i) = \bzero_d,\quad
    \hat\btheta_{\mathcal{F}}^{(k)}:\frac{1}{N}\sum_{i=1}^N \h(\btheta^{(k)};s_i^{(k)},\x_i) = \bzero_d.  
\end{equation}
Further, let $\btheta = (\btheta^{(1)^{\trans}},\btheta^{(2)^{\trans}}, \dots, \btheta^{(K)^{\trans}})^{\trans} \in\mathbb{R}^{q}$ with $q=dK$. 

In SPI, we apply an augmented estimation strategy to extract additional information of $\bbeta_0$ from surrogates. We consider the class of surrogate-powered estimators defined as 
\begin{equation*}
   \mathcal{B} = \{\hat\bbeta: \hat\bbeta = \hat\bbeta_{\mathcal{V}} - \bGamma(\hat\btheta_{\mathcal{V}} - \hat\btheta_{\mathcal{F}}), \bGamma\in\mathbb{R}^{p\times q}\},
\end{equation*}
where $\bGamma$ is a fixed and finite weighting matrix. This augmentation procedure is appealing: both $\hat\btheta_{\mathcal{V}}$ and $\hat\btheta_{\mathcal{F}}$ are consistent estimators of $\btheta_0$, ensuring the asymptotic unbiasedness of $\hat\bbeta$. Moreover, since the surrogate acts as a proxy for $y$ and shares the same covariates, $\btheta_0^{(k)}$ is likely to be informative about $\bbeta_0$. Potential variance reduction can be achieved with a specific choice of $\bGamma$. When $\bGamma$ is the zero matrix, the surrogate-powered estimator reduces to $\hat\bbeta_{\mathcal{V}}$ with no utilization of surrogates. 


Under certain conditions, the surrogate-powered estimator is asymptotically normal. We summarize the properties of the estimator in the following lemma. 
\begin{lemma}\label{lemma:asyvar}
Let $\hat\bbeta\in\mathcal{B}$ be a surrogate-powered estimator. Under regularity conditions, we have $\sqrt{n}(\hat\bbeta-\bbeta_0)\xrightarrow{d}N(\bzero, r\var(\tilde\bpsi))$ for a pair of $(\bGamma, \rho(\x, \s))$ with $r=\bbE\{\rho(\x,\s)\}$. 
The influence function for $\hat\bbeta$ is
\begin{equation*}
     \tilde\bpsi = \frac{R\bpsi}{\rho(\x, \s)} + \frac{\rho(\x, \s)-R}{\rho(\x, \s)}\bGamma\beeta,
\end{equation*}
where $\bpsi\in\mathbb{R}^p$ is the full data influence function for $\bbeta$, and $\beeta = (\beeta^{(1)^{\trans}}, \beeta^{(2)^{\trans}}, \ldots, \beeta^{(K)^{\trans}})^{\trans}\in\mathbb{R}^q$ with each $\beeta^{(k)}\in\mathbb{R}^d$ being the full data influence function for $\btheta^{(k)}$, $k=1, \ldots, K$. 
Furthermore, the asymptotic variance is  
\begin{equation}\label{eq:asvar}
    \var(\tilde\bpsi) = \bbE\Bigl\{\frac{1}{\rho(\x,\s)}\bpsi\bpsi^{\trans}\Bigr\}-\bbE\Bigl\{\frac{1-\rho(\x,\s)}{\rho(\x,\s)}(2\bpsi\beeta^{\trans}\bGamma^{\trans}-\bGamma\beeta\beeta^{\trans}\bGamma^{\trans}) \Bigr\}. 
\end{equation}
\end{lemma}
 
Based on \eqref{eq:asvar}, we can minimize the asymptotic variance with respect to $\bGamma$ and $\rho(\x,\s)$ to obtain estimators that are more efficient than $\hat\bbeta_{\mathcal{V}}$. 
However, the joint optimization can be complicated. Thus, we first consider the case when the form of $\rho(\x,\s)$ is known, minimize the asymptotic variance with respect to $\bGamma$, and propose the following \textbf{Base-SPI} estimator. 

\begin{theorem}\label{thm:basespi}
Under the same conditions of Lemma~\ref{lemma:asyvar}, the Base-SPI estimator is defined as
\begin{equation}\label{eq:spi}
    \hat\bbeta_{SPI}^* = \hat\bbeta_{\mathcal{V}} - \bGamma_0(\hat\btheta_{\mathcal{V}} - \hat\btheta_{\mathcal{F}}), 
    \quad \bGamma_0 = \bbE\Bigl\{\frac{1-\rho(\x,\s)}{\rho(\x,\s)}\bpsi\beeta^{\trans}\Bigr\}\bbE\Bigl\{\frac{1-\rho(\x,\s)}{\rho(\x,\s)}\beeta\beeta^{\trans}\Bigr\}^{-1}.
\end{equation}
The Base-SPI estimator achieves the highest efficiency among the class of surrogate-powered estimators with $\sqrt{n}(\hat\bbeta_{SPI}^*-\bbeta_0)\xrightarrow{d}N(\bzero, r\bSigma_{SPI^*})$ and 
{\footnotesize
\begin{equation}\label{eq:opt-asyvar}
    \tr\{\bSigma_{SPI^*}\}=\bbE\Bigl\{\frac{1}{\rho(\x,\s)}\bpsi^{\trans}\bpsi\Bigr\} -\tr\Bigl(\bbE\Bigl\{\frac{1-\rho(\x,\s)}{\rho(\x,\s)}\bpsi\beeta^{\trans}\Bigr\}\bbE\Bigl\{\frac{1-\rho(\x,\s)}{\rho(\x,\s)}\beeta\beeta^{\trans}\Bigr\}^{-1}\bbE\Bigl\{\frac{1-\rho(\x,\s)}{\rho(\x,\s)}\beeta\bpsi^{\trans}\Bigr\}\Bigr).
\end{equation}
}
\end{theorem}

The first term in \eqref{eq:opt-asyvar} corresponds to the variance when $\bGamma$ is the zero matrix, while the second term is the efficiency gain through surrogate integration. Consequently, the Base-SPI estimator ensures a harmless impact on the asymptotic efficiency when $\bGamma_0$ is known. 

In practice, however, the optimal Base-SPI estimator in equation~\eqref{eq:spi} is unattainable, as the matrix $\bGamma_0$ is generally unknown and must be estimated from the data. As a product of two expectations, a consistent estimate of $\bGamma_0$ can be obtained via
\begin{equation}\label{eq:gammaest1}
    \hat\bGamma_0 = \Bigl\{\sum_{i\in\mathcal{V}}\frac{1-\rho(\x_i,\s_i)}{\rho(\x_i,\s_i)^2}\hat\bpsi_i\hat\beeta_i^{\trans}\Bigr\}
    \Bigl\{\sum_{i=1}^N\frac{1-\rho(\x_i,\s_i)}{\rho(\x_i,\s_i)}\hat\beeta_i\hat\beeta_i^{\trans}\Bigr\}^{-1},
\end{equation}
or
\begin{equation}\label{eq:gammaest2}
    \hat\bGamma_0 = \Bigl\{\sum_{i\in\mathcal{V}}\frac{1-\rho(\x_i,\s_i)}{\rho(\x_i,\s_i)^2}\hat\bpsi_i\hat\beeta_i^{\trans}\Bigr\}
    \Bigl\{\sum_{i\in\mathcal{V}}\frac{1-\rho(\x_i,\s_i)}{\rho(\x_i,\s_i)^2}\hat\beeta_i\hat\beeta_i^{\trans}\Bigr\}^{-1},
\end{equation}
where $\hat\bpsi_i$ and $\hat\beeta_i$ are the estimates for $\bpsi$ and $\beeta$ obtained by plugging in $\hat\bbeta_{\mathcal{V}}$, and $\hat\btheta_{\mathcal{V}}$ or $\hat\btheta_{\mathcal{F}}$. To handle singular matrices, we add a small positive constant to the diagonal in practice. We thereby denote the practical version of Base-SPI as 
$\hat\bbeta_{SPI}=\hat\bbeta_{\mathcal{V}}-\hat\bGamma_0(\hat\btheta_{\mathcal{V}}-\hat\btheta_{\mathcal{F}})$.

\section{SPI+: Regularization with No-harm Property}
\label{sec:regaug}

While the above $\hat\bbeta_{SPI}$ appears to solve our problem, the efficiency gain may be outweighed by the error introduced from estimating the $\bGamma_0$ matrix of dimension $p\times q$ based on a finite sample, making $\hat\bbeta_{SPI}$ no longer asymptotically equivalent to $\hat\bbeta_{SPI}^*$.
We focus on the setting when $p=o(\sqrt{n})$ and $d=o(\sqrt{n})$, while allowing $q=dK$ to grow at a faster rate. 
Consider the following error decomposition of the practical Base-SPI
\begin{equation}\label{eq:est-error}
    \hat\bbeta_{SPI} = \hat\bbeta_{\mathcal{V}} - \bGamma_0(\hat\btheta_{\mathcal{V}}-\hat\btheta_{\mathcal{F}}) + (\bGamma_0 - \hat\bGamma_0)(\hat\btheta_{\mathcal{V}}-\hat\btheta_{\mathcal{F}}),
\end{equation}
where the term $(\bGamma_0 - \hat\bGamma_0)(\hat\btheta_{\mathcal{V}}-\hat\btheta_{\mathcal{F}})$ represents the estimation error induced by $\hat\bGamma_0$. Each of the quantities $\hat\bbeta_{\mathcal{V}}$, $\hat\btheta_{\mathcal{V}}$, $\hat\btheta_{\mathcal{F}}$, and $\hat\bGamma_0$ is of order $O_p(n^{-1/2})$ when obtained from \eqref{eq:est-beta}, \eqref{eq:est-theta}, and \eqref{eq:gammaest1} or \eqref{eq:gammaest2}. Consequently, the first and second terms on the right-hand-side of \eqref{eq:est-error} are of order $O_p(n^{-1/2})$, while the third error term is of order $O_p(qn^{-1})$. As the dimension of nuisance parameter increases with the inclusion of more surrogates, we allow $q$ to scale polynomial with $n$, i.e., $q = O(n^c)$ for some constant $c$. When $c\geq 0.5$, the error term becomes non-negligible, inflating both the variance and the bias of the Base-SPI estimator. 
As $c$ increases further, $\hat\bbeta_{SPI}$ eventually underperforms $\hat\bbeta_{\mathcal{V}}$ that uses no surrogates. This reversal demonstrates incorporating more surrogates in Base-SPI can harm rather than boost estimation performance without proper control on the dimension of nuisance parameters.

To address the issue stemming from $\hat\bGamma_0$, we introduce a sparsity pursuit to obtain a parsimonious estimate $\hat\bGamma_s$ of $\bGamma_0$. From \eqref{eq:gammaest2}, $\hat\bGamma_0$ can be viewed as the solution to a multi-response linear regression problem, in which $\bpsi$ is the $p$-dimensional response, and $\beeta$ is the $q$-dimensional covariates. Consequently, the dimension reduction problem in estimating $\bGamma_0$ is formulated as the following regularized multi-response regression with $p\times q$ parameters
\begin{equation}\label{eq:optim2}
    \hat\bGamma_s \in \mathop{\arg\min}_{\bGamma\in\mathbb{R}^{p\times q}}\left\{\frac{1}{2n}\sum_{i\in\mathcal{V}}\frac{1-\rho(\x_i,\s_i)}{\rho(\x_i,\s_i)^2}\|\bpsi_i - \bGamma\beeta_i \|^2 + \lambda\Omega(\bGamma)\right\},
\end{equation}
where $\lambda>0$ is the tuning parameter and $\Omega(\cdot)$ is a penalty function determined by the proposed sparsity assumption on the $\bGamma_0$ matrix. By construction, the multi-response regression problem does not have intercept. Specifically, we consider two penalty functions in this work. 
\begin{itemize}
    \item Lasso: Assume $\bGamma_0$ is element-wise sparse, we consider
    \begin{equation}\label{eq:lasso}
        \hat\bGamma_s \in \mathop{\arg\min}_{\bGamma\in\mathbb{R}^{p\times q}}\left\{\frac{1}{2n}\sum_{i\in\mathcal{V}}\frac{1-\rho(\x_i,\s_i)}{\rho(\x_i,\s_i)^2}\|\bpsi_i - \bGamma\beeta_i \|^2 + \lambda\sum_{l=1}^p\sum_{j=1}^q|t_{lj}|\right\},
    \end{equation}
    where $t_{lj}$ is $(l,j)$th element of $\bGamma$. 
    \item Group lasso: Assume $\bGamma_0$ is column-wise sparse, we consider
    \begin{equation}\label{eq:glasso}
        \hat\bGamma_s \in \mathop{\arg\min}_{\bGamma\in\mathbb{R}^{p\times q}}\left\{\frac{1}{2n}\sum_{i\in\mathcal{V}}\frac{1-\rho(\x_i,\s_i)}{\rho(\x_i,\s_i)^2}\|\bpsi_i - \bGamma\beeta_i \|^2 + 
    \lambda\sum_{j=1}^q\|\bt_j\|\right\},
    \end{equation}
    where $\bt_j$ is the $j$th column of $\bGamma$.
\end{itemize}

With the sparse estimator $\hat\bGamma_s$ for estimation error control, we now propose the \textbf{SPI+} estimator (surrogate-powered inference with regularization) in the following theorem. 
\begin{theorem}\label{thm:regularization}
We define the SPI+ estimator by
\begin{equation}
    \hat\bbeta_{SPI+} = \hat\bbeta_{\mathcal{V}} - \hat\bGamma_s(\hat\btheta_{\mathcal{V}} - \hat\btheta_{\mathcal{F}}),
\end{equation}
where $\hat\bGamma_s$ is estimated from \eqref{eq:lasso} or \eqref{eq:glasso}. Then, under certain conditions, $\hat\bbeta_{SPI+}$ is asymptotically equivalent to $\hat\bbeta_{SPI}^*$ and $\sqrt{n}(\hat\bbeta_{SPI+}-\bbeta_0)\xrightarrow{d}N(\bzero, r\bSigma_{SPI^*})$ with $\bSigma_{SPI^*}$ defined in \eqref{eq:opt-asyvar}.
\end{theorem}
Through the sparsity pursuit, the SPI+ estimator maintains asymptotic equivalence with the optimal but infeasible Base-SPI estimator, thereby inheriting its desirable no-harm property 
and effectively preventing degraded estimation performance. 
Our analysis reveals that this asymptotic equivalence relies on the error term in the decomposition~\eqref{eq:est-error} being $o_p(n^{-1/2})$, which can be achieved when $\bGamma_0$ possesses a certain sparse structure. When $\bGamma_0$ has at most $s_*$ nonzero elements, the error term is negligible if $s_*^2\log(pq)=o(n)$ under lasso regularization. Similarly, under group lasso regularization, the requirement becomes $pq_*^2(p\vee \log q)=o(n)$ where $q_*$ is the number of nonzero columns in $\bGamma_0$.
The no-harm property can still hold when $\bGamma_0$ is only weakly sparse, which does not require exactly zero entries in $\bGamma_0$, but instead allow its entries to decay to negligible magnitudes at a sufficient rate.

There is an intuitive interpretation behind the sparsity of $\bGamma_0$.
When the labeling rule $\rho(\x,\s)=r$ for all subjects, it can be shown that 
under a linear regression model, the entries of $\bGamma_0$ correspond to the partial correlations between the validated outcome $y$ and the surrogate outcome $s$ after adjusting for $\x$. Consequently, an all-zero column in $\bGamma_0$ indicates that $y$ and one specific surrogate have no remaining association conditional on all other surrogates. This type of relationship can be common and expected, as the multiple surrogates are likely to capture similar information by predicting the same underlying label $y$. 

\begin{remark}
    Beyond the error in estimating $\bGamma_0$, there may exist an additional source of error. The accuracy of the final SPI estimator also depends on the properties of $\hat\bbeta_{\mathcal{V}}$, $\hat\btheta_{\mathcal{V}}$, and $\hat\btheta_{\mathcal{F}}$. While we consider these estimators being well-defined and asymptotically unbiased in the previous discussion, small-sample bias can arise when fitting generalized linear models (GLM) on the limited validation set.
    To reduce the bias, we adopt Firth's bias-reduced regression~\citep{firth1993bias}, which corrects first-order asymptotic bias by solving a modified score equation. 
    Additionally, Firth's method provides stable estimation in the presence of separation issues under logistic regression~\citep{puhr2017firth}.
\end{remark}

\section{SPI++: Adaptivity for Cost-Effective Labeling}
\label{sec:sampling}

While Base-SPI and SPI+ derive the optimal $\bGamma$ under a predetermined, non-optimizable labeling rule, real-world annotation processes, allow one to design the optimal labeling rule before labeling begins. 
To leverage this flexibility, we further propose \textbf{SPI++}, which combines SPI+ with adaptive labeling to achieve additional efficiency gains by selecting subjects that yield a smaller asymptotic variance.

Consider $\bGamma$ as fixed. The asymptotic variance of a surrogate-powered estimator in equation~\eqref{eq:asvar} can be reformulated as 
{\footnotesize
\begin{equation*}
     \var\{\bbE(\bpsi\mid\x,\s)\} + \bbE\Bigl(\bbE\Bigl\{\frac{1}{\rho(\x,\s)}(\bpsi-\bGamma\beeta)(\bpsi-\bGamma\beeta)^{\trans}\mid\x,\s\Bigr\} - \{\bbE(\bpsi\mid\x,\s)-\bGamma\beeta\}\{\bbE(\bpsi\mid\x,\s)-\bGamma\beeta\}^{\trans}\Bigr),
\end{equation*}}
where the labeling probability $\rho(\x,\s)$ affects only the second term. Under A-optimality, the optimal labeling rule $\rho(\x,\s)$ can be obtained by solving
\begin{equation}\label{eq:optrho}
    \rho(\x,\s)= \mathop{\arg\min}_{\rho} \bbE\Bigl\{ \frac{\bbE(\|\bpsi - \bGamma\beeta\|^2\mid \x,\s)}{\rho(\x,\s)}  \Bigr\}, 
    \quad \mbox{s.t.}\ \bbE\{\rho(\x,\s)\}\leq \pi,
\end{equation}
where $0<\pi<1$ is a pre-specified constraint, representing limited validation resources. When subjects are selected independently, following  \citet{wang2023maximin}, the optimal rule is
\begin{equation}\label{eq:optimrule}
    \rho^*(\x,\s;\sigma,c^*) = \mathrm{1}\{\sigma > c^*\} + \frac{1}{c^*}\sigma\mathrm{1}\{\sigma\leq c^*\},
\end{equation}
where $\sigma = \sqrt{\bbE(\|\bpsi-\bGamma\beeta\|^2\mid
\x,\s)}$ and $c^*$ is chosen so that $\bbE\{\rho^*(\x,\s;\sigma,c^*)\}=\pi$. Solution \eqref{eq:optimrule} is essentially a truncated version of the solution obtained by directly applying the Cauchy--Schwarz inequality to problem~\eqref{eq:optrho}, ensuring that the selection probability remains in $[0,1]$. It is straightforward to verify that the proposed labeling rule is more efficient than uniform sampling when $\rho(\x,\s)=\pi$.

To obtain SPI++, we set $\bGamma=\hat\bGamma_s$ in \eqref{eq:optimrule} to combine SPI+ with the proposed labeling rule. The implementation of the labeling rule in practice requires estimating $\sigma$ for each subject. Since $\sigma$ relies on the validated outcome $y$, a small pilot sample is first obtained using uniform sampling or other non-optimal methods to collect initial labels. We then model $z=\log\{\exp(\|\bpsi-\hat\bGamma_s\beeta\|^2)-1\}$ as a function of $(\x,\s)$, and use the predictions from the fitted model to approximate $\log\{\exp(\sigma^2)-1\}$ for all unlabeled subjects. We denote $\hat\sigma$ as the prediction of $\sigma$, which is then used to determine the selection probability for labeling the remaining subjects.

While a larger pilot sample improves the accuracy of the estimated selection rule for subsequent labeling, it reduces the number of informatively labeled subjects in the final validation set under a fixed validation budget. To balance these two concerns, we propose a multiwave labeling procedure, where the labeling rule can be adaptively updated by incorporating newly validated outcomes in each wave. Assume that the pilot sample comprise a proportion $\kappa$ of the total validation set, while the remaining validation samples are equally allocated across $M$ subsequent waves, with approximately $N\{(1-\kappa)\pi/M\}$ newly-labeled subjects in each wave. In the $m$th subsequent wave, we adjust the number of already selected subjects and update the labeling rule by solving $c_m^*$ from 
\begin{equation}\label{eq:updaterule}
    \frac{1}{N}\sum_{i=1}^N(1-R_{mi})\rho^*(\x_i, 
    \s_i; \hat\sigma_{mi}, c_m^*)
    = \frac{\pi(1 - \kappa)}{M}, 
\end{equation}
where $R_{mi}=1$ if the $i$th object has been labeled before the $m$th wave and $R_{mi}=0$ otherwise, and $\hat\sigma_{mi}$ is estimated with all validated subjects before the $m$th wave. With this sequential multiwave labeling, the exact labeling probability after the $m$th subsequent wave can be computed as  
{\small
\begin{equation}\label{eq:updatep}
\begin{split}
    \tilde\rho_{mi}
    &=\kappa\pi + (1-\kappa)\pi\rho^*(\x_i,\s_i;\hat\sigma_{1i},c_1^*) + (1-\kappa)\pi\{1-\rho^*(\x_i,\s_i;\hat\sigma_{1i},c_1^*)\}\rho^*(\x_i,\s_i;\hat\sigma_{2i},c_2^*) + \dots \\
    &\mathrel{\phantom{=}}{}+ 
    (1-\kappa)\pi\prod_{j=1}^{m-1} \{1-\rho^*(\x_i,\s_i;\hat\sigma_{ji},c_j^*)\}\rho^*(\x_i,\s_i;\hat\sigma_{mi},c_{m}^*). 
\end{split}
\end{equation}}
It holds that $\bbE(\tilde\rho_{Mi})=\pi$, and $\tilde\rho_{mi}$ will be used in place of $\rho(\x_i,\s_i)$ to compute $\hat\bbeta_{\mathcal{V}}$, $\hat\btheta_{\mathcal{V}}$, and $\hat\bGamma_s$.

Combining SPI+ with this multiwave adaptive labeling procedure, we get the \textbf{SPI++} estimator, of which the detailed estimation steps are summarized in Algorithm~\ref{alg:1}. 

\begin{algorithm}[tb]
	\caption{SPI++: surrogate-powered inference with adaptive multiwave labeling}
    \label{alg:1}
    \KwIn{$\{\x_i, \s_i\},\,i=1,\ldots,N$, $\pi$, $\kappa$, $M$ }
    \textbf{Step 1}: Pilot sample labeling

    1. Select subjects for labeling with $\tilde\rho_{0i}=\pi\kappa$, and form the initial validation set $\mathcal{V}$\;
    2. Compute $\hat\btheta_{\mathcal{F}}$ and $\hat\beeta_i$ for $i=1, \ldots, N$.
    
    \textbf{Step 2}: Multiwave labeling

    \For{$1\leq m\leq M$}{
    1. Obtain $\hat\bbeta_{\mathcal{V}}$, $\hat\btheta_{\mathcal{V}}$, and $\hat\bpsi_i$ based on $i\in\mathcal{V}$ and $\tilde\rho_{(m-1)i}$ \;
    2. Get the regularized estimate $\hat\bGamma_s$ with $\tilde\rho_{(m-1)i}$\;
    3. On $\mathcal{V}$, fit $z\sim \x+\s$ with $z=\log(\exp(\|\bpsi-\hat\bGamma_s\beeta\|^2)-1)$\;
    4. Get $\hat z_i$ and $\hat\sigma_i = \sqrt{\log\{\exp(\hat z_i)+1\}}$ for $i\notin \mathcal{V}$\;
    5. Estimate the labeling rule $\rho^*(\x_i,\s_i;\hat\sigma_{mi},c_m^*)$ from \eqref{eq:updaterule}\;
    6. Conduct selection, obtain new labels and update the validation set $\mathcal{V}$ \;
    7. Update the exact labeling probability $\tilde\rho_{mi}$ by \eqref{eq:updatep}.
    }
    \textbf{end}
    
    \textbf{Step 3}: Obtain the SPI++ estimator with 
    \begin{equation*}
        \hat\bbeta_{SPI++} = \hat\bbeta_{\mathcal{V}} - \hat\bGamma_s(\hat\btheta_{\mathcal{V}} - \hat\btheta_{\mathcal{F}})
    \end{equation*}
    from the final validation set and labeling probability $\tilde\rho_{Mi}$. 

    \KwOut{$\hat\bbeta_{SPI++}$}
\end{algorithm}

\section{Simulation Study}
\label{sec:simu}

We conduct simulation studies to evaluate the performance of the SPI framework in terms of estimation efficiency. 
We model both the validated outcome and surrogates using logistic regression on the same covariates. Thus, $\bbeta_0$ and $\btheta_0^{(k)}$ have the same length with $p=d$ and $q=pK$.

\subsection{Ablation Study for SPI}

In this section, an ablation study is used to explore SPI when the validated samples are selected all at once with uniform sampling. An ablation study is a set of experiments in which components of a system are removed in order to measure the impact of these components on the performance of the system. We conduct the ablation study by progressively adding key components to the estimation method. The study begins with a label-only baseline that uses weighted logistic regression on the validation set (LO). Firth's bias reduction is applied on LO to address the small-sample bias, yielding the second method (LOF). 
Surrogates are then combined. The third method (Base-SPI(Single)) uses one randomly selected surrogate~\citep{tong2020augmented}, while the fourth method (Base-SPI(Multi)) incorporates all $K$ surrogates~\citep{lu2024leveraging}, both estimating $\bGamma_0$ from equation~\eqref{eq:gammaest1}. Finally, we combine regularization and implement SPI+ with both group lasso (SPI+(GL)) and lasso (SPI+(L1)). 
For methods requiring tuning parameter selection, the 5-fold cross-validation is used. We repeat each experiment 200 times and report the average mean squared error (MSE) for coefficient estimation over the repetitions.

We consider the following three cases: 
\begin{itemize}
    \item Case 1 ($\bar y \approx 0.2$, $p=11$): $\bbeta_0 = (-2, 0.6, 0.5, -0.7, -1, -0.5, 0.7, -0.6, 0, 0, 0)$;
    \item Case 2 ($\bar y \approx 0.1$, $p=11$, more imbalanced response compared to Case 1): \\
    $\bbeta_0 = (-3.2, 0.6, 0.5, -0.7, -1, -0.5, 0.7, -0.6, 0, 0, 0)$;
    \item Case 3 ($\bar y \approx 0.2$, $p=30$, increased number of features compared to Case 1): \\
    $\bbeta_0 = (-2.5, 0.6, 0.5, -0.7, -1, -0.5, 0.7, -0.6, 0.7, 1,-2.5, 0.4, -0.6, -0.8, 1, -0.9, 0.5,\\ -0.5, 1.2, 0.9,-0.6, 0.6, 0.5, 1.3, -1.1, 0, 0, 0, 0, 0)$.
\end{itemize}
In each case, the first element of $\bbeta_0$ is the intercept.
Case~2 mimics an imbalanced classification scenario which is commonly encountered in EHR-based studies, while Case~3 increases the number of covariates.  
For each observation, the feature vector $\x$ is independently generated from $N_{p-1}(\bzero, \bSigma)$, where the $(i,j)$th element of $\bSigma$ is set as $0.5^{|i-j|}$. The response $y$ is generated from the model $y\sim Bernoulli(\tau)$, where $\tau=\exp(\mu)/\{1+\exp(\mu)\}$ and $\mu = (1,\x)^{\trans}\bbeta_0$.

We vary the number of surrogates $K$ from 2 to 10. Each surrogate $s^{(k)}$ is generated from $y$ based on pre-specified sensitivity $P(s^{(k)}=1|y=1)$ and specificity $P(s^{(k)}=0|y=0)$. 
To have multiple surrogates with varied quality, the first surrogate is assigned a sensitivity of 90\% and a specificity of 95\%. For the $2$nd to the $(K-1)$th surrogate, both sensitivity and specificity decrease by a fixed step size of 5\%. For each $K$, the last surrogate is set as a random guess of the true label $y$, with $P(s^{(K)}=1)=0.5$. In each experiment, we generate $N=4500$ independent observations. We set $\pi=1/30$ and select the validation set via uniform sampling in a single wave, resulting in a labeled subsample of approximately 150 subjects in each experiment.

\begin{figure}[tb]
    \centering
    \includegraphics[width=0.9\linewidth]{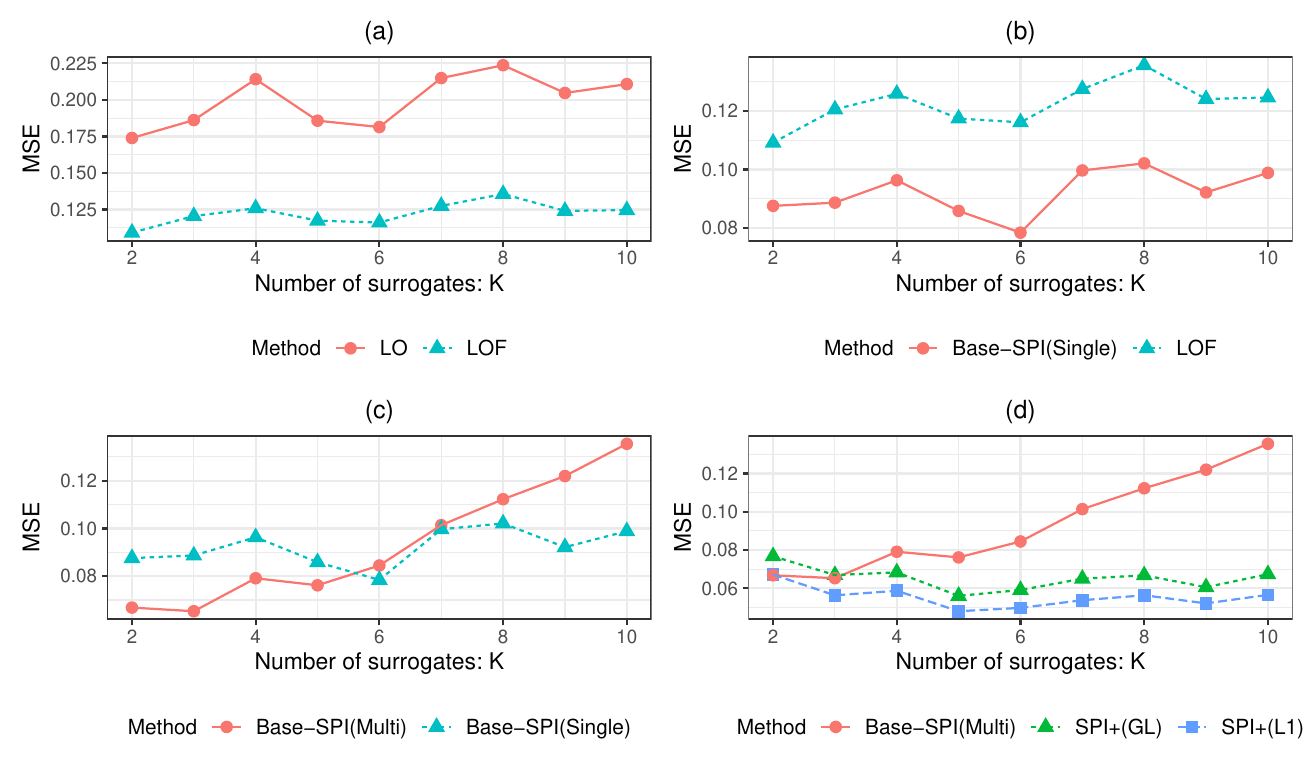}
    \caption{Simulation study: Case 1. Reported are the MSEs over 200 repetitions.}
    \label{fig:simu1}
\end{figure}

Figure~\ref{fig:simu1} presents the results under Case~1. Figures~\ref{fig:simu1}(a) and \ref{fig:simu1}(b) illustrate that applying Firth's bias reduction technique and incorporating information from the surrogate both progressively reduce the MSE. In Figure~\ref{fig:simu1}(c), Base-SPI with multiple surrogates initially reduces MSE when $K$ is small; however, as $K$ increases, the complexity of modeling surrogates gradually outweighs the efficiency gains, leading to inflated MSE. The performance of SPI+(GL) and SPI+(L1) is shown in Figure~\ref{fig:simu1}(d), which achieve the lowest MSEs when $K=4$ to $10$, attributed to the no-harm property. As for each $K$ we have a surrogate with nearly no information about $y$, the results indicate that SPI+ method is robust to highly noisy surrogates.

The results under Case~2 and Case~3 are provided in Figures~\ref{fig:simu2} and \ref{fig:simu3} in Section~\ref{supp:simu:ablation} of the Supplement. Compared with Case~1, Case~2 features a more imbalanced response. Under this scenario, SPI+ performs comparably to Base-SPI for small $K$ but outperforms it as $K$ increases. Case~3 involves a higher-dimensional predictor space with the same proportion of $y=1$ as in Case~1, and the results closely align with those observed in Case~1. These results demonstrate the robustness and efficiency of SPI+ in more challenging data settings.

Further, in each repetition, we compute standard errors for different methods and perform $z$-tests at the 5\% significance level, relying on the asymptotic normality of the estimators. 
Since the three cases include both predictive (nonzero) and non-predictive (zero) features, we evaluate the average rejection rate of the $z$-test for these two groups of features. We show the results when $K=8$ in Table~\ref{tab:testing} of the Supplement. The results indicate that Base-SPI tends to select more non-predictive features, resulting in higher false-positive rates; while Base-SPI(Multi) gives lowest rejection rates for predictive features across all three cases, providing limited power in detecting true signals. SPI+ balances true negatives and false positives,  achieving a relatively better feature selection performance than the competing methods.

\subsection{SPI++: Explore Adaptive Labeling}
\label{sec:simu-sample}

In this section, we further compare SPI+ and SPI++ when adaptive labeling can be applied. We denote the SPI++ methods with different regularization choices as SPI++(GL) and SPI++(L1). The pilot sample comprises 30\% of the total validation set, followed by four additional waves with equal sampling ratios. We use support vector regression to estimate $\sigma$, which produces fairly good performance. We repeat the experiment $200$ times and report the average MSEs. 

\begin{figure}[tb]
    \centering
    \includegraphics[width=\linewidth]{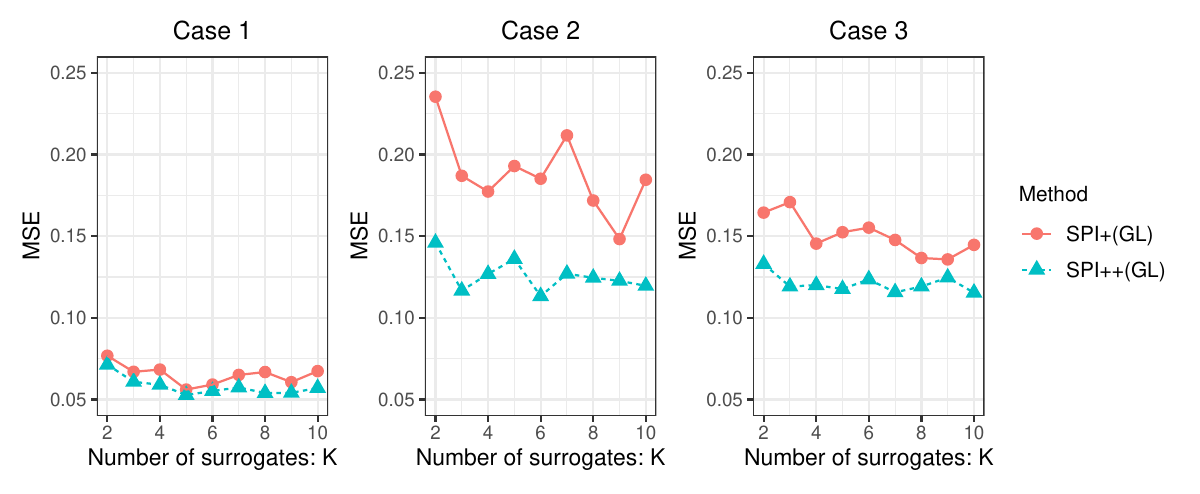}
    \caption{Simulation study: adaptive labeling with group lasso regularization. Reported are the MSEs over 200 repetitions.}
    \label{fig:sample_gl}
\end{figure}

Figure~\ref{fig:sample_gl} reports the performance of SPI+ and SPI++ with group lasso regularization. For Case~1 and Case~3, where the proportion of minority class in the response is around 20\%, adaptive labeling yields slightly better MSE compared to uniform sampling. Specifically, in Case~1, the proportion of ones in the validation sample increases by 18.8\% and 21.4\% on average under SPI++(GL) and SPI++(L1), respectively, and by 15.8\% and 13.9\% in Case~3. On the other hand, we observe substantial improvements in MSE with SPI++ under Case~2, where the response is more imbalanced. 
The efficient sampling framework reduces variance through boosting the minority class, resulting in approximately 42.3\% and 43.6\% increases in the number of the minority class in the validation sample. The results for SPI++(L1) are consistent with those for SPI++(GL). See Figure~\ref{fig:sample_l1} in Section~\ref{supp:simu:sample} of the Supplement.  

As we consider a multiwave labeling procedure, we further investigate how the MSE is affected by two factors: the total number of waves after the pilot sample labeling ($M$), and the pilot sample size ($n_0$). Simulations are conducted under the same settings as Case~2.

To evaluate the impact of the number of waves, we consider $M \in\{0, 1, 2, 3, 4, 5, 7, 10, 18, 35\}$ with $\kappa=0.3$. Here, $M=0$ corresponds to the single-wave SPI+ methods when the validated samples are selected via uniform sampling. For multiwave methods, the number of subjects sampled per wave after the pilot sampling is approximately $\{52, 35, 26, 21, 18, 13, 10, 6, 3\}$. Substantial efficiency gains are observed when transitioning from the single-wave labeling to adaptive multiwave labeling. Additional improvements are achieved as the number of subsequent  waves increases from 1 to 4. However, when the number of newly-labeled subjects per wave becomes comparable to the number of predictors $p$, the marginal reduction in MSE becomes minimal. These findings suggest that $M$ can be chosen to ensure a sufficient sample size increment in each wave, balancing computational efficiency and statistical performance. The details are reported in Figure~\ref{fig:wave} of Supplement~\ref{supp:simu:sample}.

To investigate the effect of pilot sample size, we vary the pilot sample ratio $\kappa$ from 0.2 to 0.8 in step of 0.1, with the total validation sample size of $n\approx 150$ and $M=4$. This results in $n_0$ ranging from 30 to 120. A larger pilot sample improves the accuracy of sampling rule estimation, while it reduces the number of informatively sampled subjects in the final validation set. As suggested in Figure~\ref{fig:ratio} of  Supplement~\ref{supp:simu:sample}, the MSE has an increasing trend for almost all values of $K$ when $n_0\geq 75$, suggesting a pilot ratio no greater than 0.5.

\section{Discussion}
\label{sec:diss}

In this work, we introduce the surrogate-powered inference (SPI) framework of three progressive variants, for dealing with large-scale unlabeled data accompanied by multiple surrogate labels, while leveraging a limited amount of human-validated labels. Base-SPI reduces variance through augmented estimation, while SPI+ leverages regularization to control estimation error, safely accommodating multiple surrogates of varying quality. Building on this, SPI++ further improves efficiency when the labeling process can be adaptively controlled.
By a fusion of surrogate-powered augmentation, regularization, and data-adaptive labeling, SPI effectively harnesses the benefits of data-integration  while safeguarding against potential harm to inference.
Numerical results have demonstrated that SPI can make productive use of information from surrogates. The proposed toolbox is widely applicable to different model settings and surrogate types.

There are many directions for future research. 
Given that noisy covariates are commonly encountered in medical studies~\citep{wu2024real}, it is pressing to extend SPI to handle more general settings when both surrogate outcomes and surrogate covariates are present as auxiliary variables. Moreover, we generally assume that the validated data are accurate and can serve as the ground truth. However, in some conditions, even reviewed labels may be noisy. When all available labels are potentially noisy, the problem falls into the regime of weakly supervised learning~\citep{ratner2017snorkel}. 
From a theoretical perspective, we develop SPI under the setting when $p=o(\sqrt{n})$ and $d=o(\sqrt{n})$. It is of interest to extend SPI to high-dimensional settings when both the number of covariates $p$, and the number of surrogates $d$, grow with the validation sample size $n$. 
Though \citet{hou2023surrogate} and \citet{deng2024optimal} have explored similar settings, how to ensure no-harm integration while combining adaptive labeling remains an open question.

SPI can also be explored under the multi-task learning settings, in which more than one outcome require simultaneous validation and modeling~\citep{shepherd2023multiwave}. In such scenarios, different loss functions can be considered in the labeling design by taking into account the overlapping structure between outcomes. While the current SPI framework is designed to enhance statistical inference, improving prediction performance under the no-harm and adaptivity principles can be interesting. The current SPI framework bridges data-integration and idea of active learning~\citep{zrnic2024active}. When considering prediction tasks, a shift towards combining data-integration and reinforcement learning could be helpful. 

Last but not least, we will explore tighter combining of human intelligence and machine intelligence within the SPI framework in a broader range of real applications.

\clearpage

\bibliographystyle{chicago}
\bibliography{ref}

\clearpage

\appendix

\section{Additional Simulation Results}

\subsection{Additional Ablation Study Results}
\label{supp:simu:ablation}

\begin{figure}[H]
    \centering
    \includegraphics[width=0.9\linewidth]{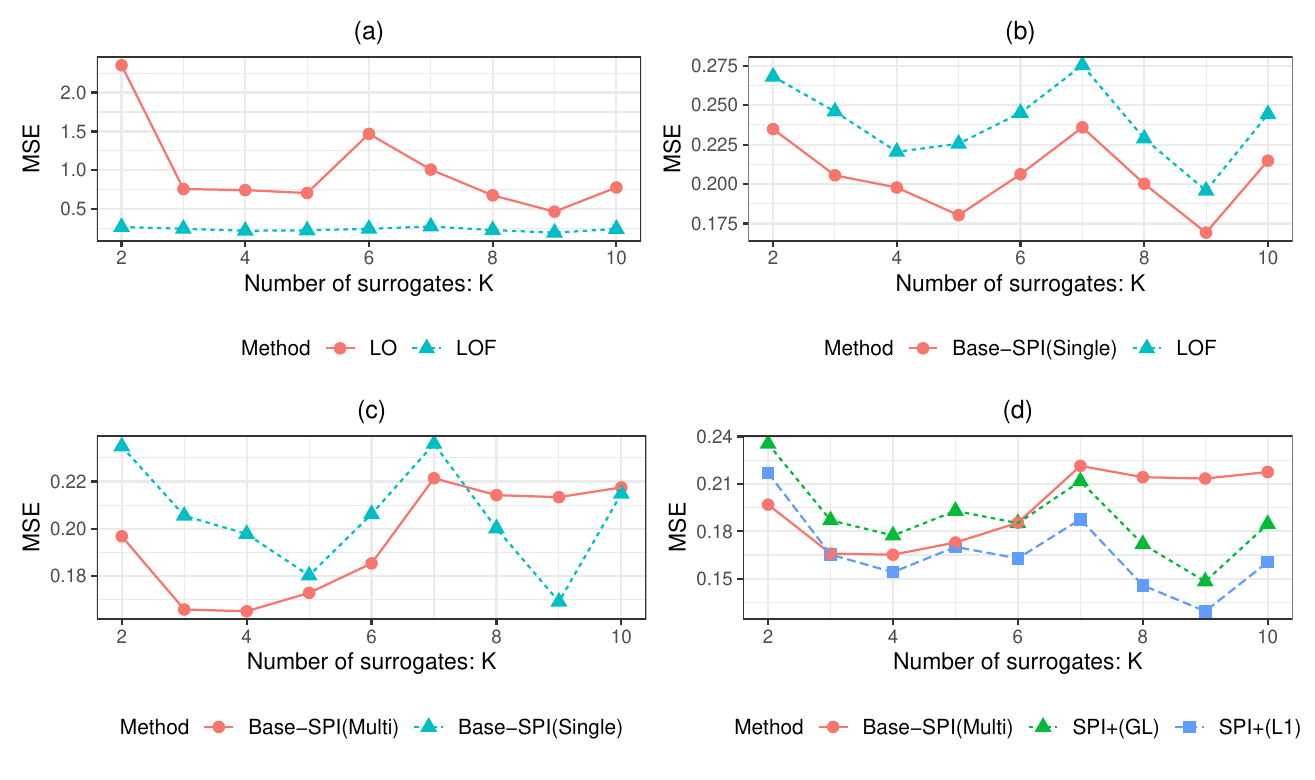}
    \caption{Simulation study: Case 2. Reported are the MSEs over 200 repetitions.}
    \label{fig:simu2}
\end{figure}

\begin{figure}[H]
    \centering
    \includegraphics[width=0.9\linewidth]{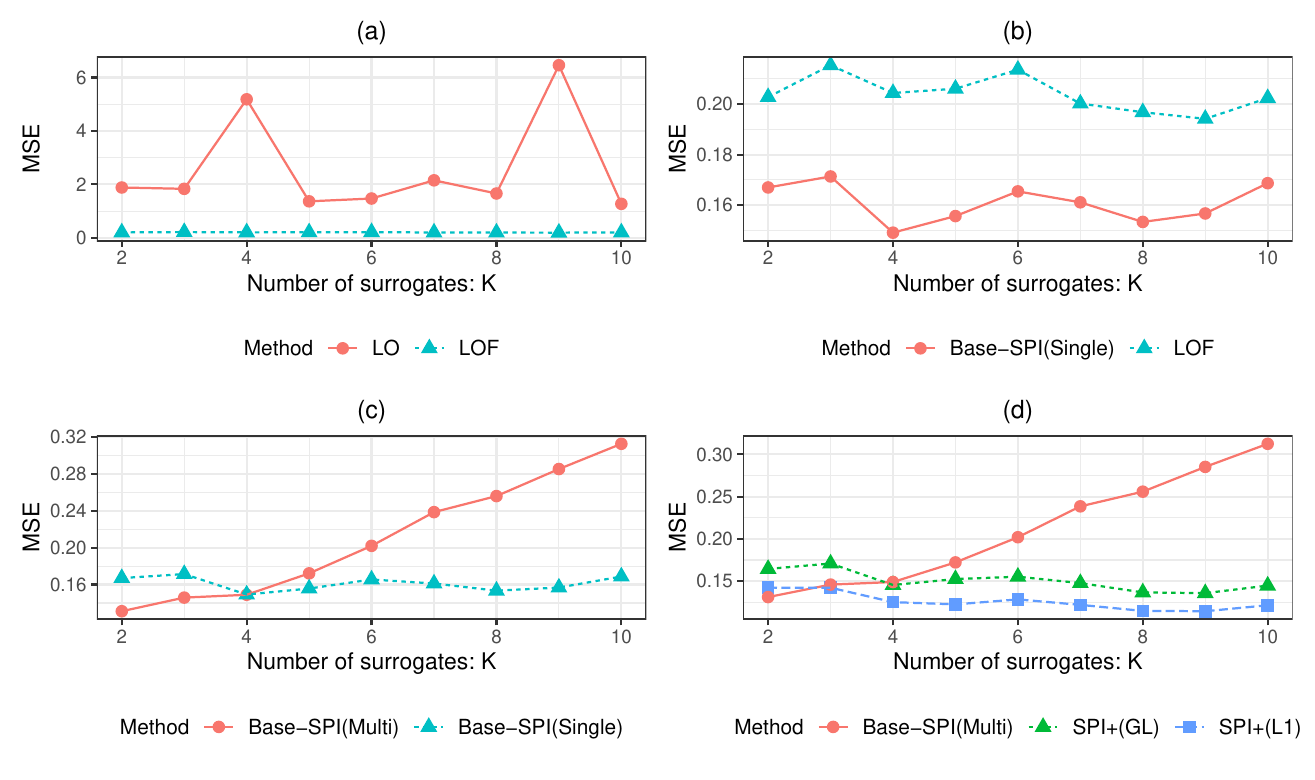}
    \caption{Simulation study: Case 3. Reported are the MSEs over 200 repetitions.}
    \label{fig:simu3}
\end{figure}

For simulated Case~2 and Case~3, we report the estimation results in Figure~\ref{fig:simu2} and Figure~\ref{fig:simu3}. Table~\ref{tab:testing} records the test rejection rates under each method over the 200 repetitions. 

\begin{table}[h]
    \centering
    \caption{Simulation study: rejection rates when $K=8$. Reported are the average rejection rates for predictive/non-predictive features over 200 repetitions.}
    \label{tab:testing}
\resizebox{0.7\textwidth}{!}
{%
    \begin{tabular}{l|ccc}
\toprule
Method  & Case 1 & Case 2 & Case 3\\
\midrule
LO & 0.649\,/\,0.113 & 0.479\,/\,0.132 & 0.622\,/\,0.158\\
LOF & 0.635\,/\,0.112 & 0.489\,/\,0.137 & 0.600\,/\,0.142\\
Base-SPI(Single) & 0.768\,/\,0.147 & 0.612\,/\,0.221 & 0.721\,/\,0.210\\
Base-SPI(Multi) & 0.543\,/\,0.113 & 0.374\,/\,0.127 & 0.252\,/\,0.060\\
SPI+(GL) & 0.913\,/\,0.223 & 0.734\,/\,0.276 & 0.830\,/\,0.277\\
SPI++(GL) & 0.931\,/\,0.188 & 0.775\,/\,0.240 & 0.831\,/\,0.233\\
SPI(L1) & 0.926\,/\,0.133 & 0.748\,/\,0.211 & 0.817\,/\,0.207\\
SPI++(L1) & 0.943\,/\,0.198 & 0.790\,/\,0.233 & 0.839\,/\,0.218\\
\bottomrule
\end{tabular}
}
\end{table}

\subsection{Results for Adaptive Labeling in Section~\ref{sec:simu-sample}}
\label{supp:simu:sample}

Figure~\ref{fig:sample_gl} compares single-wave (SPI+) and multiwave adaptive sampling (SPI++) with lasso regularization.

\begin{figure}[H]
    \centering
    \includegraphics[width=0.9\linewidth]{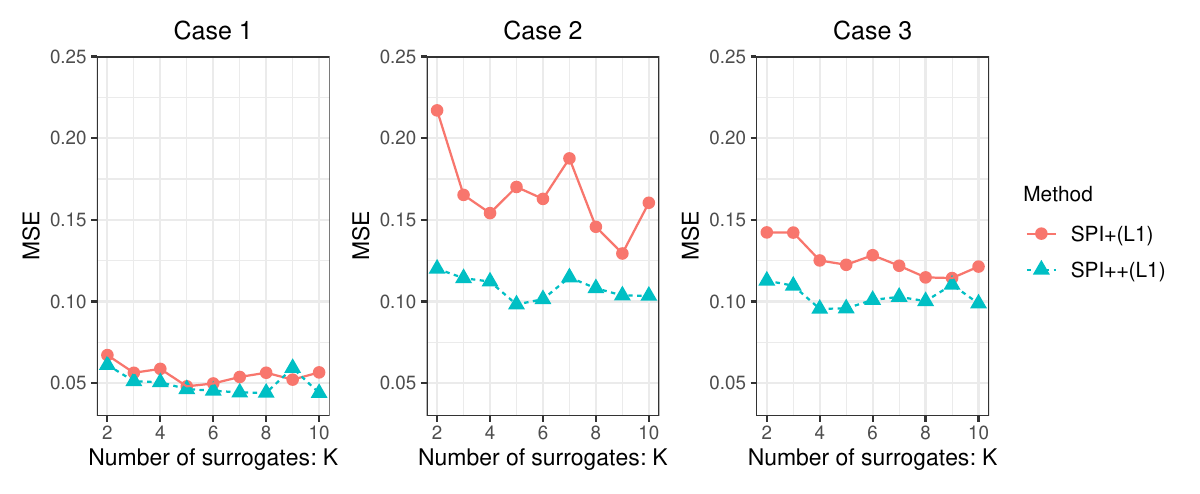}
    \caption{Simulation study: adaptive labeling with lasso regularization. Reported are the MSEs over 200 repetitions.}
    \label{fig:sample_l1}
\end{figure}

\begin{figure}[H]
    \centering
    \includegraphics[width=0.65\linewidth]{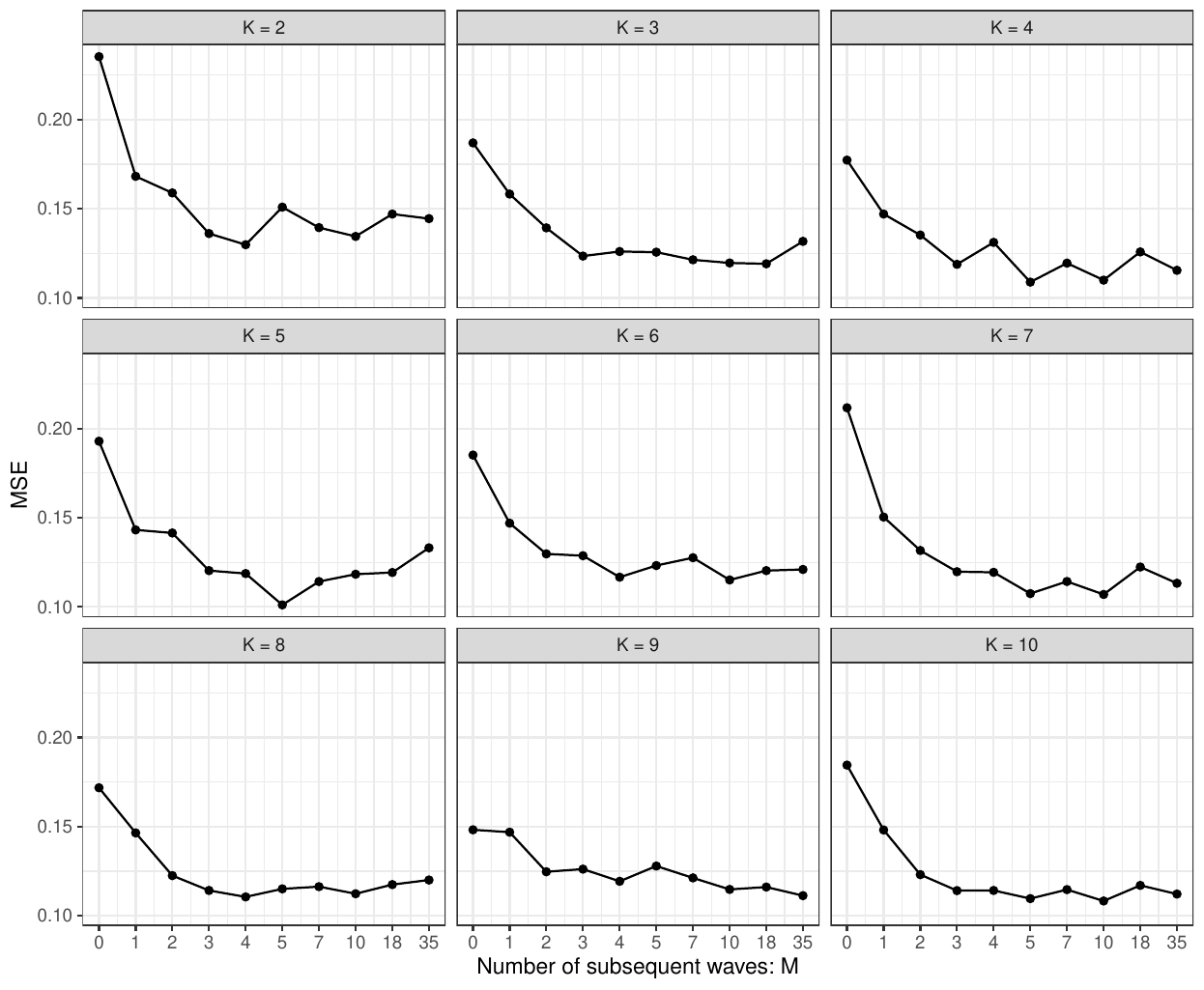}
    \caption{Simulation study: effects of different numbers of waves. $K$ denotes the number of surrogates. Reported are the MSEs over 200 repetitions.}
    \label{fig:wave}
\end{figure}

\begin{figure}[H]
    \centering
    \includegraphics[width=0.65\linewidth]{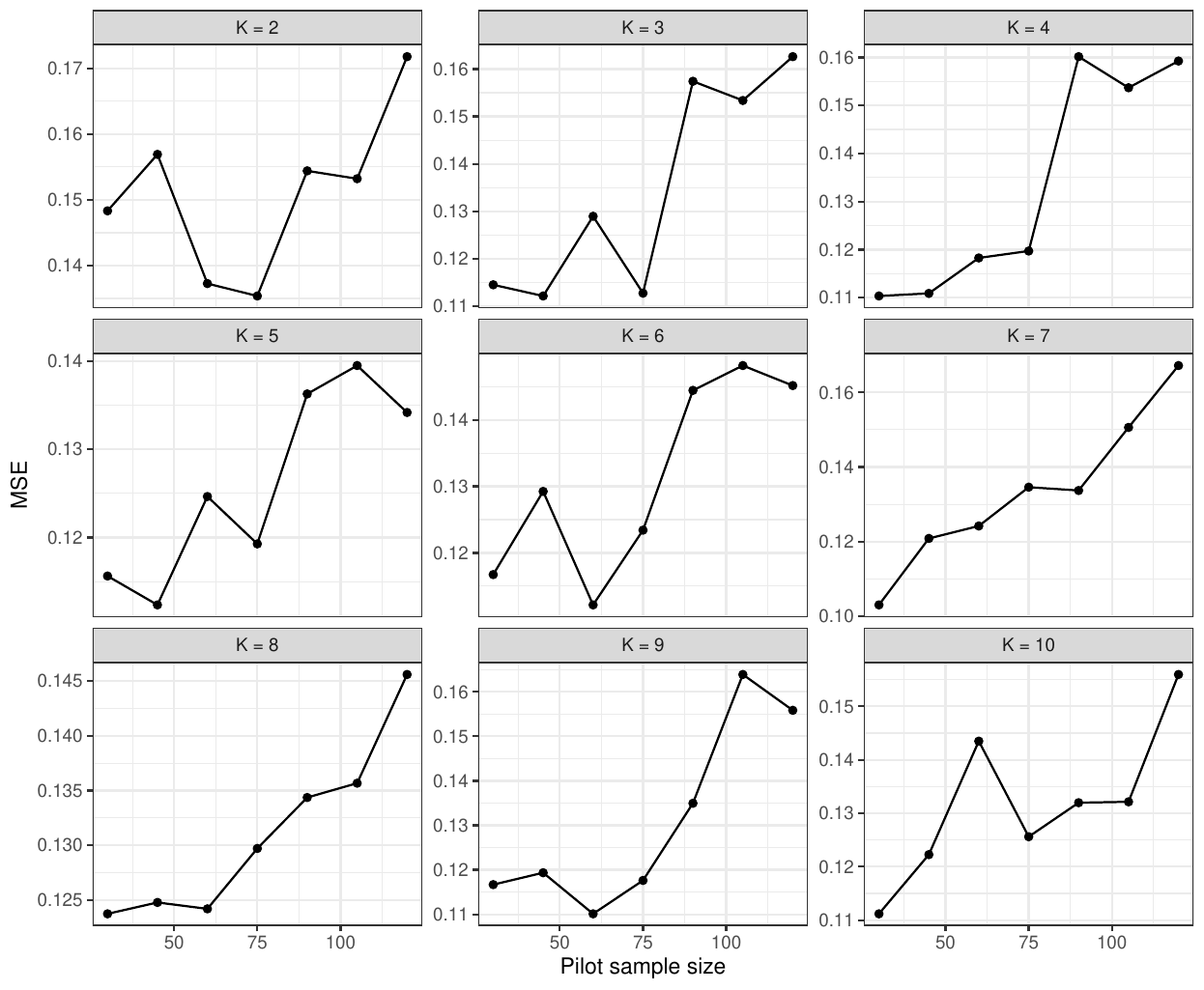}
    \caption{Simulation study: effects of different pilot sample sizes. $K$ denotes the number of surrogates. Reported are the MSEs over 200 repetitions.}
    \label{fig:ratio}
\end{figure}

Figure~\ref{fig:wave} reports the performance of SPI++(GL) for different numbers of waves, in which each subfigure represents a different number of surrogates $K$. The first point in each subfigure corresponds to the result of SPI+(GL) when a single wave with uniform sampling is applied for validation set selection. Figure~\ref{fig:ratio} reports the performance of SPI++(GL) for different pilot sample sizes. The performance of SPI++(L1) has a similar pattern to SPI++(GL).

\end{document}